\documentclass[
 amsmath,amssymb,
 aps, prx, twocolumn, superscriptaddress
]{revtex4-2}
\usepackage{xcolor}
\usepackage{graphicx}
\usepackage{float}
\usepackage{dcolumn}
\usepackage{bm}

\usepackage{mathtools}
\usepackage{amsmath,amssymb}
\usepackage{graphics,graphicx,color, calc}
\usepackage[english]{babel}
\usepackage{mathrsfs}
\usepackage[mathcal]{eucal}
\usepackage{lipsum}
\usepackage{xr}
\usepackage{stackengine}
\usepackage{booktabs}
\usepackage{soul}
\usepackage[normalem]{ulem}

\DeclareMathOperator{\argmax}{arg max}
\let\Re\relax

\DeclareMathOperator{\Re}{Re}

\DeclareMathOperator{\tr}{tr}

\begin{document}

\preprint{APS/123-QED}

\title{Nonreciprocal Model B: The role of mobilities and nonreciprocal interfacial forces}% Force line breaks with \\

\author{Bibhut Sahoo}
\affiliation{Institute for Theoretical Physics, University of G\"ottingen, 37077 G\"ottingen, Germany}

\author{Rituparno Mandal}
\affiliation{Soft Condensed Matter Group, Raman Research Institute, Bengaluru 560080, Karnataka, India}

\author{Peter Sollich}%
 \email{peter.sollich@uni-goettingen.de}
\affiliation{Institute for Theoretical Physics, University of G\"ottingen, 37077 G\"ottingen, Germany}
\affiliation{Department of Mathematics, King's College London, London WC2R 2LS, United Kingdom}

\begin{abstract}
We study a non-reciprocal version of Model B, as the continuum theory for non-reciprocal particle mixtures. In contrast to non-reciprocal Cahn-Hilliard models, it is important in this context to consider the dependence of mobility coefficients on the local concentrations. We show that a homogeneous state that is linearly stable for one form of the mobility can be unstable for a different form of mobility, an effect that would be impossible in equilibrium and implies a crucial role for mobilities in non-reciprocal mixtures. For unstable homogeneous states we study the spinodal dynamics governing the onset of phase separation. We find, again in contrast to non-reciprocal Cahn-Hilliard models, that exceptional point transitions between static and oscillatory instabilities are generically avoided by first order transitions where the spinodal lengthscale changes discontinuously. At these transitions we find intricate spinodal dynamics with two competing lengthscales, one governing a static instability and the other an oscillatory instability, {\it i.e.}\ one that generates travelling waves. We demonstrate that, depending on interaction strengths, more complex transitions can occur in the spinodal dynamics, including coexistence of three lengthscales and first order transition lines,  terminated by critical points, between distinct static instabilities.
Finally, we explore the effects of additional non-reciprocity in the interfacial chemical potentials, which would generically be expected when obtaining Model B by coarse graining from a non-reciprocal particle model. 
We show that interfacial non-reciprocity can increase the region in the spinodal phase diagram where oscillatory instabilities occur, but only up to a certain boundary that we establish analytically and demonstrate numerically.
\end{abstract}

\maketitle

\section{Introduction}
Phase separation is a ubiquitous phenomenon in nature and has been an intense topic of investigation for decades because of its broad relevance. This includes alloys~\cite{LANGER197153} and liquid mixtures~\cite{Bray01032002} in physical systems, but more recently also social phenomena like flocking~\cite{PhysRevLett.114.068101} as well as phase separation in biological systems~\cite{annurev:/content/journals/10.1146/annurev-cellbio-100913-013325}, in the latter context ranging from the formation of biomolecular condensates~\cite{condensates_challenge} to 
genome organization~\cite{gibson_organization_2019}, immune response~\cite{xiao_phase_2022, doi:10.1126/science.aat1022}, neurodegenerative diseases~\cite{zbinden_phase_2020}, SARS-Cov2 infections~\cite{ aguzzi_phase_2016} and cancer~\cite{phaseseparation_diseases}. When a homogeneous system is brought into an unstable region of its phase diagram, {\it e.g.}\ by a quench, {\it i.e.}\ a rapid reduction in temperature, fluctuations will start to grow exponentially in a process known as spinodal decomposition, leading eventually to phase separation into two or more phases. At early times in this process, fluctuations grow predominantly at a characteristic length scale. The resulting spinodal patterns initiate the formation of domains of new phases~\cite{CAHN1961795} that then grow by Ostwald ripening.

The above processes are well studied for phase separating systems in an equilibrium setting. However, many real-life systems are out of equilibrium, {\it e.g.}\ because the interactions are mediated by a non-thermal environment. The resulting effective interactions can be {\em non-reciprocal} in the sense that they violate the action-reaction symmetry, {\it{i.e.}} Newton's third law for interparticle forces~\cite{PhysRevX.5.011035,Banerjee2022}. Non-reciprocity is ubiquitous in the natural world in activator-inhibitor reactions, predator-prey dynamics, collective animal behavior and has been studied in systems such as active colloids~\cite{PhysRevLett.112.068301} and active-passive mixtures~\cite{PhysRevE.109.L062602}, as well as by extensive theoretical analysis~\cite{Vitelli_2021, Loos_2020,PhysRevE.108.064610, PhysRevE.109.L062602, mandal2024learningdynamicalbehaviorsphysical, PhysRevLett.134.117103, doi:10.1073/pnas.2412993122}. 
        
Recently, the role of non-reciprocal interaction in mixtures of two or more phase separating species has been widely studied in the setting of the Cahn-Hilliard model~\cite{PhysRevX.10.041009,PhysRevLett.131.107201,PhysRevLett.131.148301,johnsrud2025fluctuationdissipationrelationsnonreciprocal,PhysRevX.14.021014}. Non-reciprocity has been shown there to have interesting effects on the dynamics of phase separation, leading {\it e.g.} to travelling states \cite{doi:10.1073/pnas.2010318117}, suppression of coarsening \cite{PhysRevE.103.042602}, stabilization of mixtures against spinodal decomposition \cite{PhysRevLett.134.148301} and the emergence of small scale (Turing-like) instabilities \cite{10.1093/imamat/hxab026}. For our purposes it should be noted here that the Cahn-Hilliard model is typically written in a form inspired by magnetic systems. Its non-reciprocal analogue is therefore mostly written with a quartic free energy, while the mobility is taken to be constant~\cite{PhysRevX.10.041009, PhysRevE.108.064610, doi:10.1073/pnas.2010318117, PhysRevE.103.042602, 10.1093/imamat/hxab026, PhysRevLett.134.148301, PhysRevX.14.021014, saha2024phasecoexistencenonreciprocalcahnhilliard}.

Here we study the spinodal phase separation dynamics of a non-reciprocal Model B, where the free energy considered has a Flory-Huggins-like form that arises naturally by coarse graining a lattice model for particle mixtures with a solvent. This setting,
very recently studied in the context of pattern formation and wetting~\cite{ma2025wettingpatternformationnonreciprocal}, has no analogue of the Ising-like symmmetry under global order parameter sign change that is implicit in the non-reciprocal Cahn-Hilliard model. More importantly, the mobility in Model B naturally acquires a dependence on the local densities of the different particle species, and we will demonstrate that this plays a key role in the phase separation dynamics once non-reciprocity is present.

Studies of non-reciprocal Cahn-Hilliard models~\cite{PhysRevE.108.064610, PhysRevX.10.041009, PhysRevE.103.042602, doi:10.1073/pnas.2010318117, 10.1093/imamat/hxab026, PhysRevLett.134.148301, saha2024phasecoexistencenonreciprocalcahnhilliard, PhysRevX.14.021014} generally ignore the mobility, taking it as a constant factor. In Model B, however, the mobility generally has to be dependent on local mixture component concentrations~\cite{10.1063/5.0147206,Thewes_2024}.  For reciprocal (equilibrium) systems, this mobility only controls the rate at which equilibrium is reached; it does not affect the equilibrium behavior or the stability of the homogeneous mixture to fluctuations. In the non-equilibrium situation generated by non-reciprocity we will see that the form of the mobility plays an important role: it can modify not only the nature of the instability of a homogeneous state, but indeed determine whether such a state is unstable at all, thus altering the spinodal phase diagram.

In continuum or field-theoretic descriptions such as non-reciprocal Cahn-Hilliard or non-reciprocal Model B that involve local scalar density variables, but no vectorial order parameters such as local particle orientation, non-reciprocity is encoded in the form of the local chemical potentials. We can then distinguish two kinds of non-reciprocities that can arise. The first of these is bulk non-reciprocity, which captures the asymmetric influence of one species on the chemical potential of another in a bulk phase. This is typically introduced~\cite{PhysRevE.108.064610, Vitelli_2021, Loos_2020, PhysRevX.10.041009, doi:10.1073/pnas.2010318117, PhysRevE.103.042602, 10.1093/imamat/hxab026, PhysRevLett.134.148301, PhysRevX.14.021014, saha2024phasecoexistencenonreciprocalcahnhilliard} 
via a difference between corresponding off-diagonal elements of the interaction matrix, and quantified by a parameter $\alpha$. With this asymmetry the chemical potentials cannot 
be derived by taking derivatives of a free energy, thus already implying that the system is inherently out of equilibrium. In addition to bulk non-reciprocity one can then have non-reciprocity in the interfacial contributions to the chemical potentials, which depend on second (or higher) derivatives of local concentrations. This feature is typically neglected in non-reciprocal continuum models (but see~\cite{saha2024phasecoexistencenonreciprocalcahnhilliard}). We argue below, however, that interfacial non-reciprocity does arise quite generically when deriving Model B by coarse graining. The resulting contributions to the chemical potentials can, like the mobility, have significant effects on non-reciprocal phase separation dynamics as we will see.
        
We next give an overview of the structure of the paper. In section \ref{section: model_setup} 
we define the dynamics of non-reciprocal Model B dynamics and define more explicitly the two kinds of non-reciprocity mentioned above. We also provide details of the linear stability analysis on which most of our results are based. These are discussed in section \ref{section: results}. Focussing on binary mixtures, we demonstrate first that the form of the density-dependent mobility matrix affects the spinodal stability of homogeneous states. When such states are unstable, the mobility also affects the nature of this instability (static vs.\ oscillatory) and hence the qualitative features of the early-time phase separation dynamics (static spinodal patterns vs.\ travelling spinodal waves).
We determine the spinodal phase diagrams capturing the dynamics across a range of system compositions and find that in large regions of these, the eigenvalues determining the growth rate of unstable modes with wavevector $q$ change from real to complex or vice versa as $q$ increases. As a consequence, while a conventional stability analysis focusing on the behaviour for $q\to 0$ would predict exceptional point transitions between regions of static and travelling spinodal patterns~\cite{Vitelli_2021}, we find that these are generically {\em avoided} by first order transitions. At these transitions, static and travelling modes with different $q$ compete so that the spinodal patterns consist of a superposition of static and travelling contributions, with distinct associated length scales. For certain interaction parameters we also find more complex scenarios including competition between three modes that are growing equally rapidly. At this dynamical triple point multiple first order transition lines meet, which we characterize along with the associated critical points. We also demonstrate the emergence of small scale, Turing-like instabilities, which arise for unequal intraspecies interaction strengths and have also been seen in the non-reciprocal Cahn-Hilliard model~\cite{PhysRevE.103.042602}. Finally, we demonstrate that interfacial non-reciprocity can have similar effects to the mobility by changing the nature of spinodal instabilities, even if the mobility itself is taken as a trivial constant. We explore the associated spinodal phase diagrams and show that the region where travelling spinodal waves occur is in general enlarged by interfacial non-reciprocity, up to a bound that we characterize analytically. We conclude with a discussion and outlook in section \ref{section:discussion}.

\section{Setup of Model}\label{section: model_setup}

\subsection{Model definition}

The time evolution of the density fields $\phi_i$ of a phase separating mixture of $M$ components can be mathematically described by Model B~\cite{RevModPhys.49.435},
    \begin{eqnarray}
        \frac{\partial}{\partial t}{\phi}_i &=&- \nabla\cdot\bm{j}_i\label{modB_evolution}\\
        \bm{j}_i &=& -\sum_{j=1}^M L_{ij} \nabla\mu_j\label{modB_current}
    \end{eqnarray} 
Here $\bm{j}_i$ is the particle current of species $i$ while the $L_{ij}$ are the entries of a symmetric, positive definite mobility matrix that is in general dependent on all densities $\{\phi_i\}$. For equilibrium systems,
the chemical potential $\mu_j$ of species $j$ is obtained as the functional derivative of the free energy\begin{eqnarray}\label{chemical_potential_def}
        \mu_i = \frac{\delta F[\{ \phi_i \}] }{\delta\phi_i}
    \end{eqnarray} 
We will consider specifically a free energy describing the interactions between the particle species $i=1,\ldots,M$, treated at second virial level, while the solvent is assumed to interact only via volume exclusion. This free energy can also be obtained via a Flory-Huggins coarse-graining from a lattice model for particle mixtures~\cite{10.1063/1.1723621} and reads 
    \begin{eqnarray}
        F[\{ \phi_i \}] = \int d\boldsymbol{r}\left(T\sum_{i=0}^M \phi_i \log\phi_i +\frac{1}{2} \sum_{i,j=1}^M \epsilon _{ij}\phi_i \phi_j \right.\cr
        \left.+\frac{1}{2} \sum_{i,j=1}^M K_{ij} \nabla \phi_i \nabla \phi_j\right)\ .
        \label{free_en}
    \end{eqnarray}
Here $T$ is the temperature (we have set $k_{\rm B}=1$), while $\epsilon_{ij}$ and $K_{ij}$ are coefficients that control the bulk and surface interaction between species $i$ and $j$. The first term in Eq.~(\ref{free_en}) accounts for the entropy of mixing and the excluded volume interaction with the solvent. Here the solvent density is $\phi_0=1-\sum_{i=1}^M\phi_i$ 
so that the system including the solvent is incompressible.
The second term in Eq.~(\ref{free_en}) is the bulk free energy of intra- and inter-species interactions, while the last term is the interfacial free energy accounting for the cost of maintaining density gradients as they occur {\it e.g.}\ at an interface.
    
From Eq.~\eqref{chemical_potential_def}, the chemical potentials are 
\begin{eqnarray}
\mu_i &=& T\ln \phi_i -T\ln(1-\phi_0) + \mu_i^{\rm bulk} + \mu_i^{\rm interf}\label{mu_total}\\
\mu_i^{\rm bulk} &=& \sum_{j=1}^M \epsilon_{ij} \phi_j
, \quad 
\mu_i^{\rm interf} \,=\, -\sum_{j=1}^M K_{ij}\nabla^2\phi_j \label{mu_interface}   
\end{eqnarray}
where in the second line we have singled out the contributions from the bulk and interfacial interactions. Within the above (equilibrium) formalism, the phenomenon of phase separation is well understood. Starting in an unstable homogeneous state, the system phase separates into a final state consisting of different phases, thereby minimizing the free energy. Subject to the constraint of particle number conservation this leads to the equality conditions for chemical potentials and pressure and the corresponding common tangent construction~\cite{Peter_Sollich_2002, Bray01032002}.
    
The analysis of phase separation in non-equilibrium mixtures with non-reciprocal interactions between the species has come to the fore relatively recently, but has generated significant research interest~\cite{PhysRevE.108.064610, PhysRevX.10.041009, doi:10.1073/pnas.2010318117, PhysRevLett.131.107201, PhysRevLett.131.148301, PhysRevE.103.042602, 10.1093/imamat/hxab026, PhysRevLett.134.148301, johnsrud2025fluctuationdissipationrelationsnonreciprocal, PhysRevX.14.021014, saha2024phasecoexistencenonreciprocalcahnhilliard, Tucci_2024, PhysRevE.109.L062602}.
Most existing studies of non-reciprocal phase separation consider a Cahn-Hilliard free energy with non-reciprocity only in the bulk interactions terms. The corresponding contributions to the chemical potentials for a binary mixture can then be written as
    \begin{eqnarray}
        \mu_{1}^{\rm bulk}&=&\epsilon_{11}\phi_1 + (\bar{\epsilon}+\alpha)\phi_2
        \label{mubulk1}\\
        \mu_{2}^{\rm bulk}&=&\epsilon_{22}\phi_2 + (\bar{\epsilon}-\alpha)\phi_1
        \label{mubulk2}
    \end{eqnarray}
where $\alpha$ indicates the strength of the bulk non-reciprocity. For $\alpha\neq 0$ the chemical potentials can no longer be derived as derivatives of a free energy, making non-reciprocal systems non-variational in the sense of Ref.~\cite{PhysRevE.103.042602}. 

As alluded to in the interaction, if one starts from a mixture of particles with non-reciprocal pairwise interaction forces and coarse grains to obtain a continuum theory using Dean's approach~\cite{David_S_Dean_1996}, then non-reciprocity appears not only in the bulk free energy but also in the interfacial terms (see App.~\ref{appendix_deanNR}). In a binary mixture the corresponding chemical potential contributions can then be written as 
    \begin{eqnarray}
        &\mu_1^{\rm interf} = -K_{11} \nabla^2\phi_1 - (\bar{K}+\delta) \nabla^2\phi_2
        \label{interf_non_reciprocity1}
        \\
        &\mu_2^{\rm interf} =-K_{22} \nabla^2\phi_2 - (\bar{K}-\delta) \nabla^2\phi_1
\label{interf_non_reciprocity2}
    \end{eqnarray}
with $\delta$ being the strength of interfacial non-reciprocity.
    
In this paper we will focus on the interplay of the density-dependent mobility in non-reciprocal Model B with bulk non-reciprocity. For this purpose we consider two commonly studied forms of this density dependence~\cite{10.1063/5.0147206,Thewes_2024}
    \begin{eqnarray}
        L_{ij}=\phi_i\delta_{ij},~~ L_{ij}=\phi_i\delta_{ij}-\phi_i\phi_j\ .
\label{mobility_forms}
    \end{eqnarray}
The diagonal form corresponds to ideal gas limit, whereas the second form with the off diagonal elements is the case for which the system favours interdiffusion between the species. (Interpolating forms between these two extreme forms have also been suggested~\cite{10.1063/5.0147206,Thewes_2024}.) We will study specifically the influence of the choice made in Eq.~(\ref{mobility_forms}) for the early time phase separation  dynamics of non-reciprocal mixtures. After this we will turn to the effects of interfacial non-reciprocity as encoded in Eqs.~(\ref{interf_non_reciprocity1},\ref{interf_non_reciprocity2}).

\subsection{Linear stability analysis}

After a system in a spatially homogeneous state has been brought into an unstable region of its phase diagram, {\it e.g.}\ by a quench, phase separation proceeds by spinodal decomposition. In the corresponding time regime, {\it i.e.}\ before domains with well-defined interfaces are formed, the dynamics can be described by linearization in the density deviations $\delta\phi_i=\phi_i-\phi_i^0$  from the homogeneous state with ``parent'' densities $\phi_i^0$. One then finds for the dynamics of the Fourier components $\delta\phi_i(q)$, for a binary system and in vector form, 
    \begin{widetext}
    \begin{eqnarray}
\label{linearise_evolution_eqn}
        \frac{d}{dt}{\boldsymbol{\delta\phi}} &=& -q^2 \boldsymbol{L}
        \begin{pmatrix}
            \frac{T}{\phi_1^0}+\frac{T}{\phi_0^ 0}+\epsilon_{11}+q^2K_{11} &\frac{T}{\phi_0^0} +(\bar{\epsilon}+\alpha) +q^2 (\bar{K}+\delta) \\
            \frac{T}{\phi_0^0} +(\bar{\epsilon} - \alpha) + q^2(\bar{K}-\delta) & \frac{T}{\phi_2^0} +\frac{T}{\phi_0^0} + \epsilon_{22} + q^2K_{22}
        \end{pmatrix}\boldsymbol{\delta\phi}
        \label{Rmatrix_explicit}\\
        &=&-q^2\boldsymbol{L}\left( \boldsymbol{H} + q^2 \boldsymbol{K} \right) \boldsymbol{\delta\phi} = \boldsymbol{R}~ \boldsymbol{\delta\phi}
        \label{Rmatrix_compact}
    \end{eqnarray}
    \end{widetext}
where $q$ is the modulus of the Fourier wavevector and $\bm{L}$ is the mobility matrix for component densities $(\phi_1^0,\phi_2^0)$. Inside the brackets in Eq.~(\ref{Rmatrix_compact}), $\bm{H}$ is the matrix containing the bulk interaction coefficients $\epsilon_{ij}$ as well as the contribution from the excluded volume interaction, while $\bm{K}$ contains the interfacial coefficients. In writing the explicit form (\ref{Rmatrix_explicit}) we have already inserted the parametrizations from Eqs.~(\ref{mubulk1}
--\ref{interf_non_reciprocity2}).

The nature of the spinodal dynamics is primarily determined by the eigenvalues of the rate matrix $\boldsymbol{R}$, which are given by
    \begin{eqnarray}
        \lambda_{1,2} = \frac{\tr\boldsymbol{R}\pm \sqrt{(\tr\boldsymbol{R})^2-4|\boldsymbol{R}|}}{2}.
        \label{lambda1_2}
    \end{eqnarray}
in terms of its trace and determinant. All phase diagrams presented below pertain to this spinodal dynamics. Understanding the late-stage dynamics analytically is rather more challenging; we therefore only give an overview of some of the non-trivial phenomenology that is visible in numerical simulations (see Sec.~\ref{subsection:long time}
).

\section{Results}\label{section: results}

\subsection{Effect of mobility on global stability}

We focus first on the interplay of bulk non-reciprocity (controlled by $\alpha$) and a density dependent mobility (\ref{mobility_forms}). Beyond overall stability, one question of interest will be the emergence of travelling spinodal waves rather than static spinodal patterns. It has been well documented that such waves can arise from bulk non-reciprocity in Cahn-Hilliard models~\cite{PhysRevX.10.041009, doi:10.1073/pnas.2010318117, PhysRevX.14.021014}, but as we will see the inclusion of a density-dependent mobility can significantly change the picture.

\begin{figure}
    \centering
        \centering
        \includegraphics[width=\linewidth]{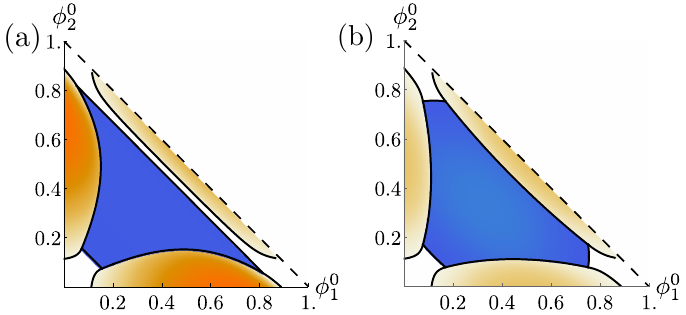}
        \caption{Spinodal phase diagrams for $q\to0$, indicating stability of homogeneous states with the given densities $\phi_1\equiv \phi_1^0$, $\phi_2\equiv \phi_2^0$.
        (a) Mobility $L_{ij}=\phi_i\delta_{ij}$, (b) mobility $L_{ij}=\phi_i\delta_{ij}-\phi_i\phi_j$. 
        In white regions the homogeneous state is stable to small density fluctuations.
        Orange regions indicate a static instability (real $\lambda_{1/2}$), blue regions a traveling instability (complex $\lambda_{1/2}$). 
        Parameter values: $T=0.1$, $\epsilon_{11}=\epsilon_{22}= -1, \epsilon_{12}=\bar{\epsilon}+\alpha=0, \epsilon_{21}=\bar{\epsilon}- \alpha=-1,K_{ij} =\delta_{ij}$.}
        \label{fig:fig1_mobility_change}
\end{figure}

Considering first the global stability of homogeneous states, which is determined by the sign of the real part of $\lambda_{1/2}$, we show in Fig.~\ref{fig:fig1_mobility_change} phase diagrams delineating the spinodal phase behavior for $q\to 0$, for the two forms of the mobility in Eq.~(\ref{mobility_forms}). One clearly observes qualitative difference between the two diagrams. Key is the observation that for the mobility $L_{ij}=\phi_i\delta_{ij}$ there is a region of stability, shown white in Fig.~\ref{fig:fig1_mobility_change}(a), in the regime of dense homogeneous states, {\it i.e.}\ large $\phi_1^0+\phi_2^0$. This region is absent in Fig.~\ref{fig:fig1_mobility_change}(b). The implication is that the kinetics of phase separation, as encoded in the mobility matrix $\bm{L}$, can determine whether a non-reciprocal system is unstable or not. This striking effect would be impossible in equilibrium, where stability is determined by properties of the free energy alone, without reference to the kinetics.

In regions of instability in the $q\to 0$ spinodal phase diagrams of Fig.~\ref{fig:fig1_mobility_change} we observe static instabilities (real $\lambda_{1/2}$, orange) as well as oscillatory ones (complex $\lambda_{1/2}$, blue) that correspond to travelling spinodal waves. One can check that the transitions between these regions occur, as in the case of non-reciprocal Cahn-Hilliard models~\cite{PhysRevE.108.064610, PhysRevX.10.041009, saha2024phasecoexistencenonreciprocalcahnhilliard, Tucci_2024, PhysRevX.14.021014}, via exceptional point transitions. At these exceptional points~\cite{Vitelli_2021} two distinct real eigenvalues merge and then become complex. The corresponding eigenvectors collapse at the exceptional point, {\it i.e.}\ become parallel to each other. We do not explore these exceptional points further because, as we will see next, the presence of a non-trivial mobility means that the nature of any instabilities cannot be predicted by considering the limit $q\to 0$ of small wavevectors. 

\begin{figure*}
    \centering
    \includegraphics{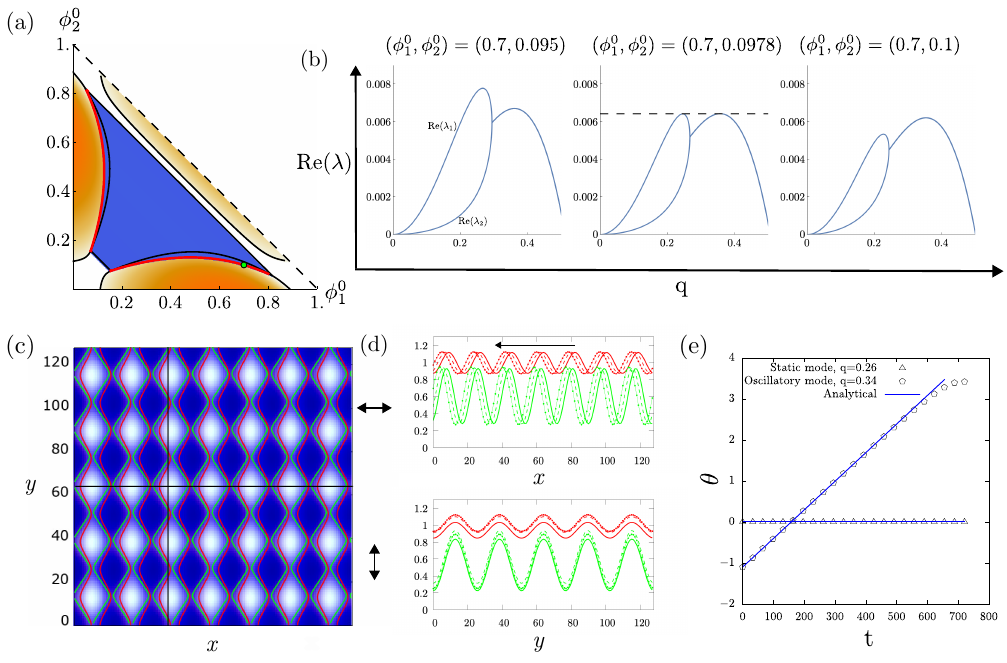}
    \caption{Exceptional points are avoided by first order transitions.
(a) Spinodal phase diagram determined from the rate matrix eigenvalues $\lambda_{1/2}$ at the physically relevant wavevector $q_{\rm max}$, at the same parameters as in Fig.~\ref{fig:fig1_mobility_change}(a). Black lines: phase boundaries as expected from spinodal analysis for $q\to0$ modes; red lines: boundary between static and oscillatory instabilities at the physical $q_{\rm max}$ (b) $\Re(\lambda_{1/2})$ vs $q$ for a series of three homogeneous state compositions with fixed $\phi_1^0=0.7$, crossing the red transition line at the green point in the vertical direction. 
The transition is first order, with a discontinuous change from a static (real eigenvalue) instability with $q^{\rm r}_{\rm max}\approx 0.26$ to an oscillatory one (complex eigenvalue) with $q^{\rm c}_{\rm max}\approx 0.34$. Middle plot: coexistence between  the two modes at the transition (green dot in (a)). (c) RGB plot of system at transition (green dot in (a)), with modulations corresponding to dominant oscillatory mode in $x$-direction and dominant real mode in $y$-direction; blue indicates dominance of species 2 ($\phi_2>\phi_2^0$), light colours indicate density below average ($\phi_1+\phi_2<\phi_1^0+\phi_2^0$). Green/red lines: contours of constant $\phi_1$ and $\phi_2$, indicating asymmetry of composition profile in $x$-direction. 
(d) Density profiles along cross-sections in the $x$-direction (top) and $y$-direction (bottom) as indicated by black lines in (c). Plotted are normalized deviations $\delta\phi_1$ (green) and $\delta\phi_2$ (red), for three different times $t=0$ (solid), $t=98.3$ (dashed), $t=196.6$ (dotted). The spinodal pattern is travelling only in the $x$-direction (top) but not the $y$-direction, as expected from the initial modulation. (e) The oscillatory growing mode in (c,d) gives for the Fourier transform $\bm{\delta\phi}(q)$ a sum of two contributions with complex phases. Isolating the one with increasing phase angle $\theta$ (App.~\ref{appendix: num_method_oscillatory_analysis}), we find $\theta$ varying linearly in time (pentagons: numerical solution of Model B, blue line: analytically predicted linear dependence). The static mode (triangles) has constant phase angle. 
}
\label{fig:fig2_firstorder_symmetriceps}
\end{figure*}

\subsection{First order transitions between static and oscillatory instabilities}\label{results: first_order_transition_symmetric_eps}

A linear stability analysis of the type above, where only small wavevectors $q$ are considered explicitly, was used also in early studies of non-reciprocal Cahn-Hilliard models~\cite{PhysRevX.10.041009, doi:10.1073/pnas.2010318117, PhysRevLett.134.148301}. This approach is valid in non-reciprocal Model B when, for example, $L_{ij}\propto K_{ij}\propto\delta_{ij}$. The interfacial $O(q^4)$ contribution to the rate matrix $\bm{R}$ is then a multiple of the identity matrix, shifting both eigenvalues $\lambda_{1/2}$ by the corresponding real value. This implies that the nature of the eigenvalues (real or complex) does not change with $q$ and can be deduced from the limit $q\to 0$.
In general, however,
\iffalse 
In general, however, the interfacial contribution to $\bm{R}$ does not commute with the leading $O(q^2)$ term so that an analysis for small $q$ no longer suffices and 
\fi 
finite wavevectors $q$ have to be studied~\cite{PhysRevLett.131.107201, PhysRevE.103.042602, 10.1093/imamat/hxab026, PhysRevLett.134.018303, PhysRevX.14.021014}.

In the case of non-reciprocal Model B considered here, the interfacial contributions to the rate matrix $\bm{R}$ will generically have a non-trivial matrix structure arising from the mobility matrix $\bm{L}$, even when the interfacial terms remains simple ($K_{ij}\propto \delta_{ij}$). For these cases, the eigenvalues $\lambda_{1/2}$ can change from real to complex or vice versa as the wavevector $q$ is varied. The nature of any spinodal instability thus has to be determined explicitly at the wavevector $q$ that is growing most rapidly and hence dominates the spinodal decomposition:
\begin{eqnarray}
    q_{\rm max} &=& \argmax_q \lambda_r(q)
\\
\lambda_r(q)&=&\max
    \{\Re(\lambda_1(q)),\Re(\lambda_2(q))\}
\end{eqnarray}

Fig.~\ref{fig:fig2_firstorder_symmetriceps}(a) shows the outcome of such an analysis, for the same parameters as in Fig.~\ref{fig:fig1_mobility_change}(a). The overall region of instability in the spinodal phase diagram has not changed, but the region of oscillatory rather than static instabilities has become larger, with the transition between these two types of instability shifted to the red line in Fig.~\ref{fig:fig2_firstorder_symmetriceps}(a). To understand the nature of the transition, we consider three compositions of the homogeneous system lying across the transition line in the vertical direction at the 
green point in Fig.~\ref{fig:fig2_firstorder_symmetriceps}(a). Fig.~\ref{fig:fig2_firstorder_symmetriceps}(b) displays the real parts of the eigenvalues $\lambda_{1/2}$ as a function of wavevector modulus $q$. Where these real parts differ, the eigenvalues themselves are real; when the real parts coincide, the eigenvalues are complex and conjugates of each other, $\lambda_2=\lambda_1^*$. 
We observe in Fig.~\ref{fig:fig2_firstorder_symmetriceps}(b) that the transition is {\em first order}: the fastest growing mode that dominates the spinodal decomposition changes discontinuously, from $q_{\rm max}\approx 0.26$ and an associated real eigenvalue of the rate matrix, to $q_{\rm max}\approx 0.34$ with a complex eigenvalue. An exceptional point lying between these two values of $q$, which in other models is frequently found to govern the transition~\cite{PhysRevE.108.064610, Vitelli_2021, PhysRevX.10.041009, doi:10.1073/pnas.2010318117, PhysRevX.14.021014} is thus {\em avoided}. Instead we have at the transition two competing, equally rapidly growing spinodal modes  (Fig.~\ref{fig:fig2_firstorder_symmetriceps}(b) middle), one generating a static pattern and the other an oscillatory one.

The above analysis indicates at the first order transition a non-trivial spinodal dynamics with instabilities of different character occurring simultaneously in the system, at two distinct length scales. To demonstrate this, we solve the non-reciprocal Model B equations (\ref{modB_evolution},\ref{mu_total},\ref{mubulk1},\ref{mubulk2}) numerically in two dimensions for an initial condition corresponding to the green dot in Fig.~\ref{fig:fig2_firstorder_symmetriceps}(a).
The initial state is taken as spatially homogeneous but with added small periodic modulations corresponding to the two fastest growing modes (Fig.~\ref{fig:fig2_firstorder_symmetriceps}(b) middle), the complex one in the $x$-direction and the real one in the $y$-direction.
Fig.~\ref{fig:fig2_firstorder_symmetriceps}(c) shows a snapshot of the configuration. Following the evolution in time we observe, as predicted, travelling waves in the $x$-direction while in the $y$-direction the pattern remains static, see Fig.~\ref{fig:fig2_firstorder_symmetriceps}(d). For ease of visualization we have factored out the exponential growth in time of the density fluctuations, {\it i.e.}\ we plot the normalized density fluctuations 
\begin{equation}
\delta{\tilde\phi}_i=\delta\phi_i/(3\sigma)\ ,
\label{sigmatilde_def}
\end{equation} where $\sigma^2=\langle(\delta\phi_1)^2+(\delta\phi_2)^2\rangle$ is the time-dependent variance of the two species densities across the system.

\begin{figure*}
    \centering
    \includegraphics{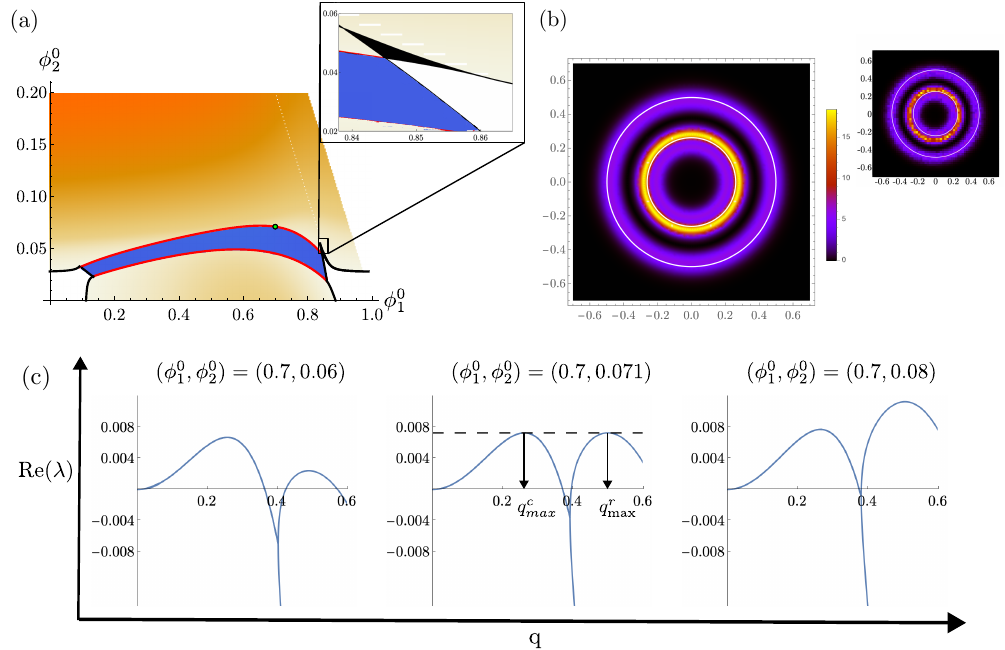}
\caption{(a) Spinodal phase diagram with $\epsilon_{22}
\neq \epsilon_{11}$; colour code as in Fig.~\ref{fig:fig2_firstorder_symmetriceps}. Inset: Zoom on region where Turing patterns occur (black area). (b) Normalized structure factor $S(q_x,q_y,t)$ vs.\ $q_x$, $q_y$ at a first order transition (green dot in (a)), for a time $t$ in the spinodal regime. {\em Two} spinodal rings are visible, at radii $q$ consistent with theoretical predictions (white circles, travelling mode: $q_{\text{max}}^\text{c}=0.2615$, static mode $q_{\text{max}}^\text{r} =0.499$). Large plot: results from linearized dynamics~\eqref{linearise_evolution_eqn}; smaller plot top right: results from direct numerical solution of Model B for system size $L=256$. (c) Real parts of rate matrix eigenvalues against wavevector $q$, for three points in phase diagram (a) vertically across phase boundary at green dot ($\phi_1^0=0.7$, $\phi_2^0$ as shown). In contrast to Fig.~\ref{fig:fig2_firstorder_symmetriceps}, the eigenvalues are complex at small $q$ and real at large $q$. Parameters as in Fig.~\ref{fig:fig2_firstorder_symmetriceps} except for $\epsilon_{22} =-3.67$, $K_{ij}=4\delta_{ij}$.
}
\label{fig:fig3_firstorder_asymeps}
\end{figure*}

The oscillatory nature of the dominant complex mode can also be confirmed from the time evolution of the Fourier transform $\bm{\delta\phi}(q)$ of the two species densities. This is a sum of two terms proportional to the two right eigenvectors of $\bm{R}(q)$, each containing a phase factor $e^{\pm i\omega t}$. If we project $\bm{\delta\phi}(q)$ onto the first contribution, its phase angle $\theta$ therefore has to vary linearly in time. This is confirmed within the spinodal time scale (until $t\approx 500$) in Fig.~\ref{fig:fig2_firstorder_symmetriceps}(e).

\subsection{First order transition in the nature of instability with asymmetric $\epsilon$}

We next show that rather more complex spinodal phase diagram topologies are possible for interaction matrices with unequal intraspecies interactions,  $\epsilon_{11}\neq \epsilon_{22}$. In Fig.~\ref{fig:fig3_firstorder_asymeps}(a) we show the spinodal phase diagram for the case $\epsilon_{22}=-3.67$. All other parameters are as in Figs.~\ref{fig:fig1_mobility_change}(a) and~\ref{fig:fig2_firstorder_symmetriceps}, with the exception of the interfacial coefficients, which we increase to $K_{ij}=4\delta_{ij}$ to keep the characteristic wavevectors of the same order as before. As in Fig.~\ref{fig:fig2_firstorder_symmetriceps}
we observe first order transitions between static and travelling instabilities (red lines in Fig.~\ref{fig:fig3_firstorder_asymeps}). Plots of the real parts of the rate matrix eigenvalues against $q$ are shown Fig.~\ref{fig:fig3_firstorder_asymeps}(c) for three points across one of the transition lines.
In contrast to Fig.~\ref{fig:fig2_firstorder_symmetriceps}(b), here the eigenvalues change from complex to real with growing $q$, rather than the other way around. Additionally, the black region in the inset of Fig.~\ref{fig:fig3_firstorder_asymeps}(a) indicates where the linear stability analysis predicts a Turing instability. The real parts of the rate matrix eigenvalues for this region are non-negative only above some non-zero threshold for the wavevector $q$. Hence, fluctuations corresponding (only) to small length scales are excited, suggesting the existence of Turing patterns. Turing patterns have also been observed in  non-reciprocal Cahn-Hilliard models~\cite{10.1093/imamat/hxab026}.

We finally illustrate for the current set of parameters an experimentally observable consequence of the first order nature transition in the spinodal dynamics: at the transition, {\em two} distinct spinodal rings can be seen in the structure factor, one each corresponding to the wavevectors of the modes dominating the static and travelling spinodal patterns, respectively. We demonstrate this effect in  Fig.~\ref{fig:fig3_firstorder_asymeps}(b),
both by evaluating the linearized theory (Eq.~\eqref{linearise_evolution_eqn}) and by a numerical solution of the non-reciprocal Model B equations. In both cases we start from an initial condition with small, spatially uncorrelated fluctuations in both densities $\phi_i$ around the homogeneous state. Specifically we plot in Fig.~\ref{fig:fig3_firstorder_asymeps}(b) the structure factor (with $q=(q_x^2+q_y^2)^{1/2}$) 
\begin{eqnarray}
    S(q_x,q_y,t) \equiv S(q,t) \propto
    \langle |\bm{\delta\phi}(q,t)|^2 \rangle
    \label{Sq_def}
\end{eqnarray}
against $(q_x,q_y)$. The proportionality factor in (\ref{Sq_def}) is chosen so that the average over all $(q_x,q_y)$ of the structure factor is unity. Note that the two-ring pattern in Fig.~\ref{fig:fig3_firstorder_asymeps} is generic in the sense that it will occur at all first order transitions in the spinodal phase diagram, including {\it e.g.}\ the ones in Fig.~\ref{fig:fig2_firstorder_symmetriceps}. 
\begin{figure*}
    \centering
    \includegraphics{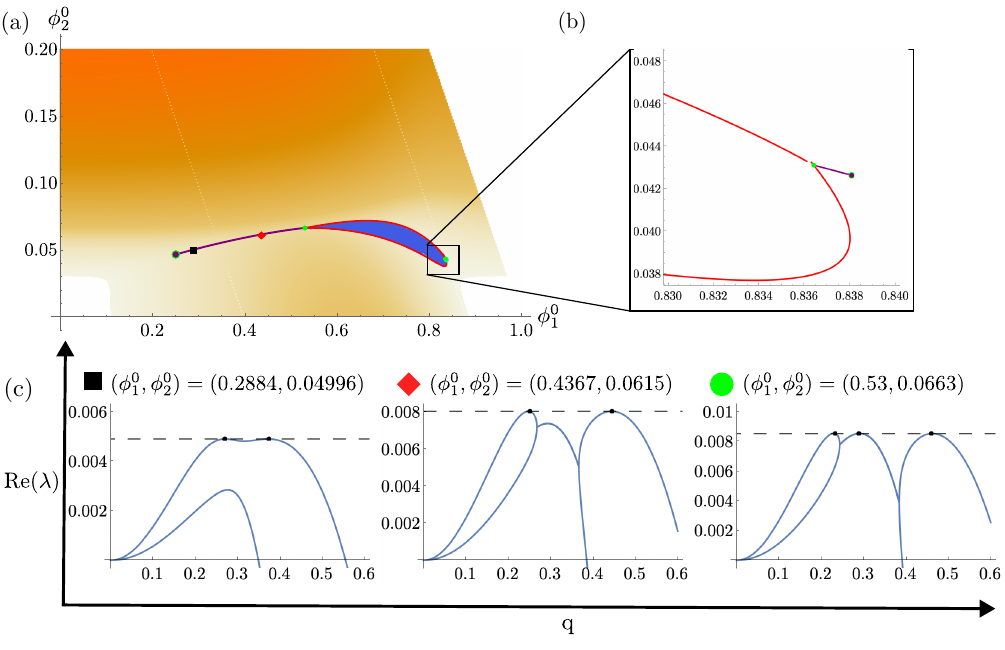}
    \caption{Non-reciprocal Model B can show coexistence of three distinct dominant spinodal modes. (a,b) Spinodal phase diagram for system considered in Sec.~\ref{sec:triple_point}. Red lines: first order transitions between dominant static and oscillatory modes; purple lines: first order transitions between two distinct static modes; green points: triple points with three competing dominant spinodal modes; purple circles with green boundary: critical points at end of static-static transition lines. (c) Real parts of rate matrix eigenvalues against wavevector $q$, at three points in the phase diagram (indicated by the symbols) along the static-static first order transition line in the left half of (a). Note the coexistence between two similar static modes when close to the critical point (left), the appearance of a subdominant oscillatory mode when moving closer to the triple point (middle), and the coexistence of three modes (two static, one oscillatory) at the triple point itself (right).}
    \label{fig:fig4_threepeaks}
\end{figure*}

\subsection{Competition of three spinodal modes}
\label{sec:triple_point}

We next demonstrate that yet further topologies 
of the spinodal phase diagram are possible in non-reciprocal Model B depending on the interaction parameters, revealing in particular homogeneous states for which the initial phase separation dynamics exhibits three competing spinodal lengthscales. 

Consider {\it e.g.}\ a system with two-species interaction matrix

\begin{equation}
\begin{pmatrix}
        \epsilon_{11} & \epsilon_{12} \\ \epsilon_{21} & \epsilon_{22}
    \end{pmatrix}
= -\begin{pmatrix}
        1 & 1/6 \\5/6 & 3.37
    \end{pmatrix}
\end{equation}
and all other parameters as in the previous subsection. The resulting spinodal phase diagram, displayed in Fig.~\ref{fig:fig4_threepeaks}(a,b), features 
several branches of first order transitions between instabilities with different lengthscales. There are in addition two triple points, where three of these lines meet and consequently three spinodal modes compete, one travelling and two static, all growing exponentially with the same $\lambda_r$ but with different associated wavevectors $q$. For the triple point at $\phi_1^0\approx 0.53$ (left green dot) in Fig.~\ref{fig:fig4_threepeaks}(a), the corresponding rate matrix spectrum is shown in Fig.~\ref{fig:fig4_threepeaks}(c) and demonstrates the coexistence of three distinct dominant spinodal modes, two of them static and one oscillatory. 

A further new feature apparent from Fig.~\ref{fig:fig4_threepeaks}(a,b) is that, from each triple point, we have one transition line emanating that indicates a first order transition between two distinct {\em static} spinodal modes. Each such static-static first order transition line ends in a critical point, indicated by a purple circle  in Fig.~\ref{fig:fig4_threepeaks}(a,b). The corresponding wavevector dependence of the rate matrix eigenvalues is displayed in Fig.~\ref{fig:fig4_threepeaks}(c) for three points along one of the static-static transition lines. Note finally that, related to the existence of the two triple points, the region in the spinodal phase diagram where travelling patterns appear (marked blue in Fig.~\ref{fig:fig4_threepeaks}(a,b)) lies entirely within the instability area, {\it i.e.}\ does not border the region where homogeneous states are stable.

\subsection{Effect of interfacial non-reciprocity on spinodal behavior}\label{results_surfNR}

So far we have investigated parameter settings in non-reciprocal Model B where the interfacial contributions to the chemical potentials in Eqs.~(\ref{interf_non_reciprocity1},\ref{interf_non_reciprocity2}) remain reciprocal ($\bar{K}=\delta=0$). However, as argued in App.~\ref{appendix_deanNR}, interfacial non-reciprocity is generically expected for continuum models of particle systems with non-reciprocal ({\it e.g.}\ pairwise) forces. We therefore next consider this case. 

To understand the effects of interfacial non-reciprocity we return to the rate matrix~\eqref{linearise_evolution_eqn} for the linearised dynamics,
\begin{eqnarray}
\boldsymbol{R}(q) = -q^2\boldsymbol{L}(\boldsymbol{H} +q^2 \boldsymbol{K} ) = -q^2 \tilde{\boldsymbol{H}} - q^4 \tilde{\boldsymbol{K}}
\label{R_explicit}
\end{eqnarray}
Here we have abbreviated by $\tilde{\boldsymbol{H}}=\bm{L}\bm{H}$ and $\tilde{\boldsymbol{K}}=\bm{L}\bm{K}$ the overall contribution coming from bulk and interfacial terms, respectively. Asking now specifically when travelling spinodal patterns arise, we require the rate matrix to have complex eigenvalues with positive real part. From (\ref{lambda1_2}) one sees that this requirement translates to
    \begin{eqnarray}
        (\tr\boldsymbol{R})^2-4|\boldsymbol{R}|<0 \quad \mbox{and}\quad \tr\boldsymbol{R}>0
    \end{eqnarray}
These conditions can be written explicitly as 
\begin{widetext}
\begin{eqnarray}
&&\left[ (\tilde{H}_{11}-\tilde{H}_{22})+q^2(\tilde{K}_{11}-\tilde{K}_{22}) \right]^2+4 (\tilde{H}_{12}+\tilde{K}_{12}q^2)(\tilde{H}_{21}+\tilde{K}_{21}q^2) <0\,,
\label{R_cond1}\\
&&\tilde{H}_{11}+\tilde{H}_{22}<-q^2(\tilde{K}_{11}+\tilde{K}_{22}).
\label{R_cond2}
    \end{eqnarray}
    \end{widetext}
The question is now when these conditions can be satisfied by tuning $\tilde{\bm{K}}$. Here we have to bear in mind that for the continuum model to make sense, there must be no instabilities for $q\to\infty$, which from~\eqref{R_explicit} means that all eigenvalues of $\tilde{\bm{K}}$ need to have positive real parts. This constraint is equivalent to 
\begin{equation}
    \tr\tilde{\bm{K}}>0 \quad \mbox{and} \quad |\tilde{\bm{K}}|>0
    \label{K_constraint}
\end{equation}
and the equivalence applies whether or not the eigenvalues of $\tilde{\bm{K}}$ are real or complex, as one can confirm from the generic expression~\eqref{lambda1_2} for the eigenvalues of $2\times 2$ matrices.

The determinant constraint in~\eqref{K_constraint} can always be satisfied by making $\tilde{K}_{12}$ and $\tilde{K}_{21}$ large and of opposite sign. The first condition~\eqref{R_cond1} for the occurrence of travelling spinodal patterns is then automatically satisfied because the second term becomes negative and large, dominating over the first term. But given the trace constraint on $\tilde{\bm{K}}$ in~\eqref{K_constraint}, the second condition~\eqref{R_cond2} always has a negative right hand side. Thus, a {\em necessary} condition for travelling spinodal patterns is that
\begin{equation}
    \tr\tilde{\bm{H}}=\tilde{H}_{11}+\tilde{H}_{22}<0
    \label{travelling_necessary}
\end{equation}
When this condition is satisfied, travelling spinodal modes always exist for some choice of $\tilde{\bm{K}}$ respecting~\eqref{K_constraint}, for wavevectors $q<(-\tr\tilde{\bm{H}}/\tr\tilde{\bm{K}})^{1/2}$ as can be seen from~\eqref{R_cond2}. 

\begin{figure}
\centering
\includegraphics[width=\linewidth]{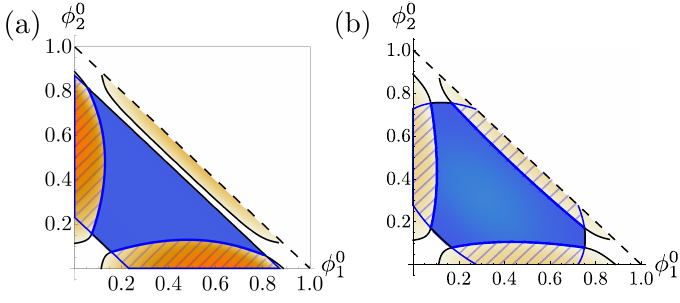}
\caption{Interfacial non-reciprocity favours  travelling spinodal patterns. Shown are the phase diagrams of Fig.~\ref{fig:fig1_mobility_change} for the same system parameters and the same mobilities: (a) $L_{ij}=\phi_i\delta_{ij}$, (b) $L_{ij}=\phi_i\delta_{ij}-\phi_i \phi_j$. The blue hatching indicates the areas where for trivial interfacial coefficients ($K_{ij}\propto\delta_{ij}$) the dominant spinodal instability is static, but where this instability {\em can} become oscillatory for sufficiently strong non-reciprocity in $K_{ij}$.}
        \label{fig:fig5_surfNR}
    \end{figure}

We show the areas in the spinodal phase diagram where~\eqref{travelling_necessary} holds by blue hatching in Fig.~\ref{fig:fig5_surfNR}, for the same system parameters as in Fig.~\ref{fig:fig1_mobility_change}. One observes that, like the overall spinodal phase diagram, also the areas where travelling spinodal waves can occur in the presence of interfacial non-reciprocity depend crucially on the form of the mobility. Of course, as the condition~\eqref{travelling_necessary} is only a necessary one, then for a given choice of $\bm{K}$ and the resulting $\tilde{\bm{K}}$, the region in the phase diagram where the dominant spinodal modes are oscillatory, {\it i.e.}\ produce travelling patterns, will generically be smaller than the hatched areas. We illustrate in App.~\ref{appendix:specific_interfacialNR_phasediagram}, however, that for suitable choices of $\bm{K}$ the difference can be made rather small.

\subsection{Long time behavior with bulk NR}\label{subsection:long time}

So far we have focussed on the phase separation dynamics in non-reciprocal Model B in the spinodal time regime, where the dynamics can be described by linearization around the initial homogeneous state. Making analytical predictions about the long-time fate of the system is a more serious challenge~\cite{PhysRevResearch.7.013234, PhysRevLett.131.148301, PhysRevLett.134.018303}. We therefore use numerical simulations of the non-reciprocal Model B equations to illustrate some of the fascinating long-time behaviors that can emerge, leaving a comprehensive study for future work.

\subsubsection{Coexistence of stable and transient domains}

\begin{figure*}
        \centering     \includegraphics[width=\linewidth]{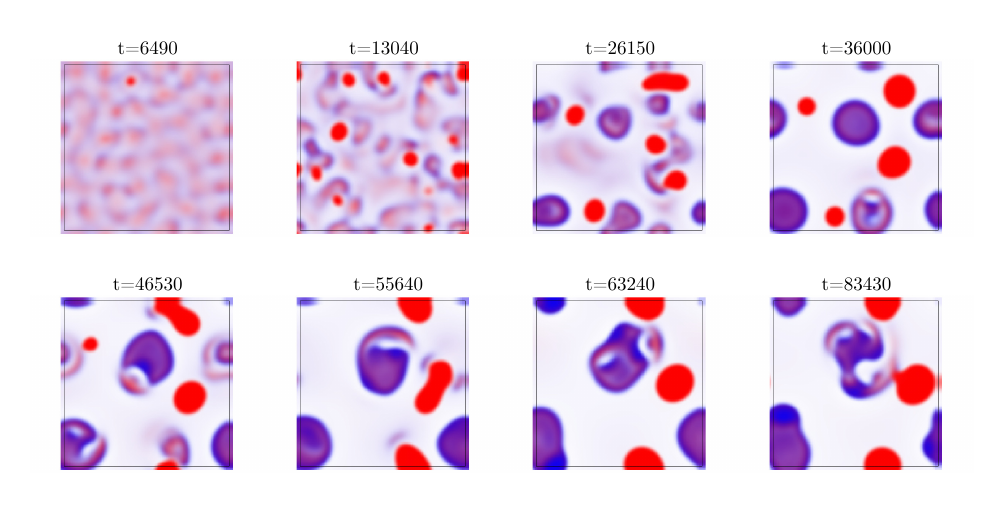}
\caption{Long-time behavior for system parameters as in Fig.~\ref{fig:fig1_mobility_change} and initial homogeneous composition $(\phi_1^0,\phi_2^0) = (0.15,0.15)$. The later snapshots show formation of stable domains of one phase (red, consisting mainly of particles of species 1) and transient domains of a second phase (blue/purple, mainly species 2). The latter keep breaking up after being ``invaded'' by the first phase.}
\label{fig:fig6_longtime_stabletransient}
    \end{figure*}

We simulate the system starting from a homogeneous state with parent density $(\phi^0_1,\phi^0_2)= (0.15,0.15)$, with added small fluctuations (see App.~\ref{appendix: numerical_scheme}). The system parameters are the same as in Fig.~\ref{fig:fig1_mobility_change}(a). At long times, well beyond the spinodal regime, we observe the appearance of  patterns where one phase forms one or several stable domains; in Fig.~\ref{fig:fig6_longtime_stabletransient} this phase contains mainly particles of species 1 so appears red. A second phase also occurs, containing predominantly particles of species 2 (blue/purple in Fig.~\ref{fig:fig6_longtime_stabletransient}), but the domains where this is found keep breaking apart as is visible in the series of snapshots in  Fig.~\ref{fig:fig6_longtime_stabletransient}. Similar patterns have recently been observed in non-reciprocal Cahn-Hilliard models~\cite{PhysRevLett.134.018303}. Intuitively, species 1 (red) ``chases'' species 2, causing it to invade domains rich in species 2 (blue), while species 2 ``runs away'' from this invasion leading to the break-up of the blue domains. Species 1 then moves back to its original (red) domain to aggregate there, allowing domains enriched in species 2 to form again so that the process can repeat.

\subsubsection{Long time chase-and-run behavior along interfaces}

\begin{figure*}
        \centering
        \includegraphics[width=1.\linewidth]{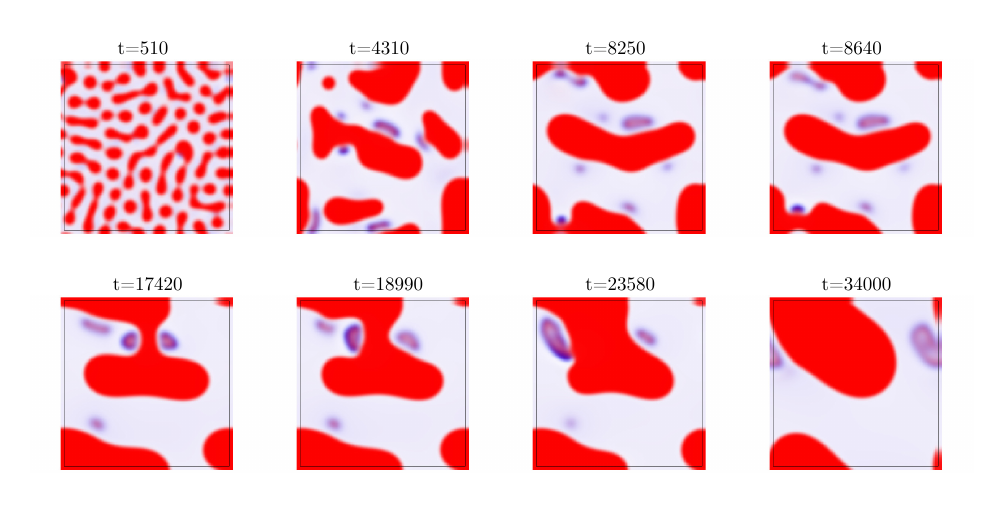}
\caption{Long-time behavior for system parameters as in Fig.~\ref{fig:fig1_mobility_change} and homogeneous parent composition $(\phi_1^0,\phi^2_0)=(0.05,0.4)$. A single domain of mainly species 1 forms (red); species 2 particles are concentrated in domains moving along the {\em interface} of the species 1 domain and locally deforming its interface.}
        \label{fig:fig7_longtime_chaserun}
    \end{figure*}

With the same parameters as above but a different initial composition of $(\phi_1^0,\phi_2^0)\equiv (0.05,0.4)$, our analysis correctly predicts that, initially, static spinodal patterns appear. These look essentially indistinguishable from those in standard reciprocal Model B (Fig.~\ref{fig:fig7_longtime_chaserun}). However, at late times, the (bulk) non-reciprocity becomes manifest as can be seen from Fig.~\ref{fig:fig7_longtime_chaserun}. 
We observe again a single domain containing predominantly species 1 particles (red). Particles of species 2 are now predominantly found in small domains that exhibit chase-and-run dynamics along the {\em interface} of the species 1 domain, reminiscent of behavior analyzed in~\cite{PhysRevE.101.022414}. The effect of species 1 wanting to ``chase'' species 2 here leads to the interface of species 1 domains (red) being locally deformed when a domain rich in species 2 (blue) is close, which in turn causes the species 2 domain to ``run'' away along the interface.

\section{Discussion}\label{section:discussion}

We have explored non-reciprocal Model B, the appropriate continuum theory for non-reciprocal particle mixtures. We have demonstrated in particular that the non-trivial density dependence of the mobility in Model B, along with interfacial non-reciprocity that we argue has to be present quite generically, can cause significant differences in the phase separation dynamics compared to {\em e.g.} non-reciprocal Cahn-Hilliard models. 
    
We have focused mostly on the spinodal dynamics after a quench into a region of the phase diagram where a spatially homogeneous system is unstable, and have analysed the corresponding spinodal phase diagrams. Our key conclusions are as follows. Firstly, the form of the density-dependent mobility is crucial in non-reciprocal Model B. In contrast to equilibrium systems, the mobility can determine whether or not a homogeneous state is stable to small density fluctuations; when it is unstable, the mobility can also determine whether this instability produces static spinodal patterns or travelling ones. The latter arise from dominant spinodal modes that have associated complex eigenvalues and therefore an oscillatory time dependence.

While in the spinodal phase diagrams of non-reciprocal models~\cite{Vitelli_2021, kreienkamp2025synchronizationexceptionalpointsnonreciprocal, PhysRevB.97.014428} including non-reciprocal Cahn-Hilliard models~\cite{PhysRevE.108.064610, PhysRevX.10.041009, PhysRevX.14.021014, saha2024phasecoexistencenonreciprocalcahnhilliard, Tucci_2024}, the transitions between static and travelling instabilities generically occur at exceptional points, we showed that in non-reciprocal Model B these exceptional points are generally avoided by first order transitions.
Crossing these first order transitions, the length scale of the spinodal patterns changes discontinuously. 
Directly at the first order transitions, two dominant spinodal modes with distinct lengthscales compete, one static and one oscillatory, resulting in intricate spinodal dynamics where travelling patterns are superimposed onto static ones. We demonstrated these effects in a system initialized with the two relevant fluctuation lengthscales in orthogonal spatial directions. For random initial density fluctuations, we showed that the same physics shows up in the structure factor, accessible {\em e.g.} via scattering experiments, 
as a double spinodal ring. 
    
We demonstrated further that more complex spinodal phase diagram topologies can result depending on the specific form of the non-reciprocal bulk interactions. These include first order transition lines between two static spinodal patterns of different lengthscales, which are terminated by critical points. Three first order transition lines can also meet, resulting in a triple point where three lengthscales compete in the spinodal dynamics.

We argued based on a simple coarse-graining of the non-reciprocal Dean equation that {\em interfacial} non-reciprocity should be generic when coarse-graining non-reciprocal particle mixtures. We showed that such interfacial non-reciprocity can increase the region in the spinodal phase diagram where travelling spinodal patterns occur, up to a simple bound that we derived analytically and confirmed numerically.

Finally we provided numerical evidence for non-trivial phenomena that arise in the long-time behaviour of non-reciprocal Model B, in particular coexistence of stable and transient domains as well as chase-and-run dynamics along domain interfaces.

In future work on the spinodal dynamics of non-reciprocal Model B, it will be interesting to explore comprehensively whether yet more complex phase diagrams can occur, with possibly even more than three competing spinodal lengthscales. Also the effects of more general forms of the Model B mobility~\cite{Thewes_2024, 10.1063/5.0147206} will be interesting to study. Beyond this, the long-time behaviour of non-reciprocal Model B is of obvious interest. One would in particular like to understand for which parameter setting the steady states can be mapped to those of appropriate non-reciprocal Cahn-Hilliard models, and conversely where novel classes of steady state behaviour appear.

\appendix

\section{Numerical methods}
\label{appendix_methods}

\subsection{First order transitions in spinodal phase diagrams}

Figs.~\ref{fig:fig2_firstorder_symmetriceps}(a), \ref{fig:fig3_firstorder_asymeps}(a) and \ref{fig:fig4_threepeaks}(a) show first order transitions within the spinodal diagrams (red lines), where two spinodal modes with different $q$ compete because their growth is governed by rate matrix eigenvalues with identical real part $\lambda_r(q)$. To locate these transitions precisely, we impose the required condition that $\lambda_r$ has two distinct maxima at wavevectors $q_1$ and $q_2$ at equal height, {\em i.e.}\ 
    \begin{eqnarray}
    \frac{d\lambda_r(q_1)}{dq} =
    \frac{d\lambda_r(q_2)}{dq} = 0 {\rm~~ and }~~ \lambda_r(q_1) = \lambda_r(q_2).
    \end{eqnarray}
We fix $\phi_1^0$ and then solve these three conditions numerically for $q_1$, $q_2$ and $\phi_2^0$ to locate the first order transition, and then vary $\phi_1^0$. One can of course conversely fix $\phi_2^0$ and solve for $q_1$, $q_2$, $\phi_1^0$; we use whichever is more convenient numerically.

\subsection{Analysis of oscillatory spinodal modes}\label{appendix: num_method_oscillatory_analysis}

To obtain the phase angle, $\theta$, of an oscillatory density Fourier mode, we proceed as follows. For fixed wavevector $q$, the solution of the linearized dynamics~\eqref{Rmatrix_compact} for density Fourier mode $\bm{\delta\phi}$ is 
\begin{eqnarray}
    \bm{\delta\phi}(t) = e^{\bm{R}t}\bm{\delta\phi}(0)=\sum_{a=1}^M e^{\lambda_a t}\bm{r}_a\bm{l}_a^{\rm T}\bm{\delta\phi}(0)
\end{eqnarray}
where $a$ labels the eigenvalues $\lambda_a$ and right and left eigenvectors $\bm{r}_a$, $\bm{l}_a$ of the rate matrix $\bm{R}$. For $M=2$ species and complex conjugate eigenvalues $\lambda_{1,2}=\lambda_r \pm i\omega$ one obtains two contributions with increasing and decreasing complex phases. Using the bi-orthogonality $\bm{l}_a^{\rm T}\bm{r}_b=\delta_{ab}$ of the eigenvectors, we can isolate each component by taking the inner product with the corresponding left eigenvector, {\it e.g.}
\begin{eqnarray}
    \boldsymbol{l}_1^{\rm T}\boldsymbol{\phi}(t) &=& A\,e^{\lambda_1 t}
    \ = \ |A|\,e^{\lambda_r t} e^{i(\varphi+\omega t)}
\end{eqnarray}
where $A=\bm{l}_1^{\rm T}\bm{\delta\phi}(0)=|A|e^{i\varphi}$. Defining $\hat{\phi}_r$ and $\hat{\phi}_i$ as the real and imaginary parts of $\boldsymbol{l}_1^{\rm T}\boldsymbol{\phi}(t)$, respectively, the phase angle $\theta=\varphi+\omega t$ can then be extracted as
\begin{equation}
    \theta=\arctan( \hat{\phi}_i/\hat{\phi}_r)
\end{equation}
The slope of $\theta$ versus $t$ then gives the imaginary part $\omega$ of the eigenvalue $\lambda_1$, {\it i.e.}\ the oscillation frequency of the mode. The corresponding (phase) velocity of the travelling spinodal wave is $\omega/q$. For static spinodal modes, the same procedure gives a constant phase angle $\theta=\varphi$, corresponding to $\omega=0$.

\subsection{Numerical solution of non-reciprocal Model B}\label{appendix: numerical_scheme}

For numerical simulations, we consider a two-dimensional square domain with periodic boundary conditions, discretized into 256$\times$256 grid points. To evaluate the spatial derivatives appearing in the Model B evolution equation, we use a Fourier pseudo-spectral scheme. For time stepping we use a semi-implicit Euler step, which contains a correction term of $\mathcal{O}(q^4)$ in the denominator in the update step~\cite{C8SM02045K, PhysRevE.60.3564}. The system is initialized with a homogeneous parent density $(\phi_1^0,\phi_2^0)$ with small random (independent Gaussian) fluctuations added at every grid point.

For the case with interfacial non-reciprocity, we needed to use adaptive time stepping. To do this, we perform an update with time step $\Delta t$ and compare with the result of two successive updates with time step $\Delta t/2$. The deviation between the two results is measured as the mean-squared deviation across the grid. If the deviation is above a set tolerance, we reduce $\Delta t$ for the next time step. This optimizes computational effort while maintaining a desired level of accuracy via the chosen error tolerance~\cite{press1996numerical}.

\subsection{Colour code for numerical simulation results}

In snapshots of the system state resulting from the numerical simulations described above we show the local composition by the RGB colour code $(\text{RGB})=
255(\phi_1, \phi_0, \phi_2)$ where $\phi_0=1-\phi_1-\phi_2$ is the vacancy density. Areas with high density of species 1 or 2 thus appear red or blue, respectively.

To make the spinodal patterns visible in Fig.~\ref{fig:fig2_firstorder_symmetriceps}(c), we rescale the deviations from the homogeneous state
as defined in~\eqref{sigmatilde_def} and show the corresponding densities $\tilde{\phi}_i = \text{squash}(\phi_i^0+\delta\tilde{\phi}_i)$. 
Here the squashing function ${\rm squash} (x)=\min(\max(x,0),1)$ restricts the values to the physical range (0,1).

\section{Spinodal phase diagram for mobility $L_{ij}=\phi_i\delta_{ij}-\phi_i\phi_j$}

\subsection{First order transitions}

\begin{figure}[H]
        \centering
        \includegraphics[width=\linewidth]{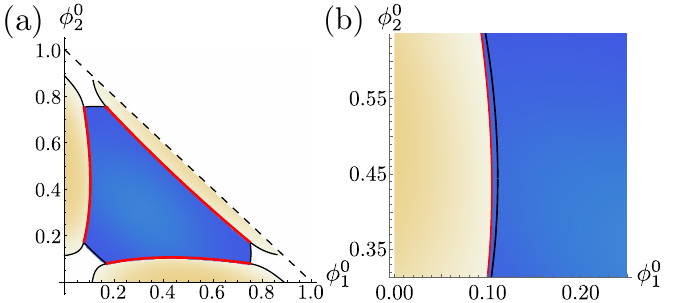}

\caption{(a) Full (finite $q$) spinodal phase diagram for system parameters as in Fig.~\ref{fig:fig1_mobility_change}(b), {\it i.e.}\ analog of Fig.~\ref{fig:fig2_firstorder_symmetriceps}(a) but with mobility $L_{ij}=\phi_i\delta_{ij}-\phi_i\phi_j$. As there, we find that exceptional point transitions (black lines between blue and yellow regions) are avoided by first order transitions (red lines). 
(b) Zoom on the transition line close to the $\phi_2^0$-axis, to show the (small) separation between the two types of transition line.}
        \label{fig:figA1}
    \end{figure}

In section ~\ref{results: first_order_transition_symmetric_eps} we demonstrated that, for non-reciprocal Model B with mobility $L_{ij}=\phi_i\delta_{ij}$, the exceptional point transitions predicted by an analysis for $q\to 0$ are avoided by first order transitions when the full $q$-dependence of the rate matrix eigenvalues is taken into account. Fig.~\ref{fig:figA1} demonstrates that the same behaviour is observed for the mobility $L_{ij}=\phi_i\delta_{ij}-\phi_i\phi_j$, with the only modification that the quantitative differences between the two types of transition line are now smaller.
    
\subsection{Effect of mobility on nature of dominant spinodal modes} 

\begin{figure}[H]
        \centering
        \includegraphics[width=\linewidth]{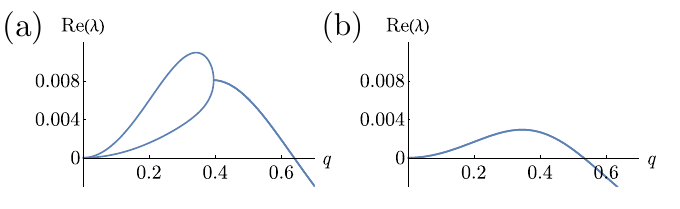}
\caption{Rate matrix spectra $\Re(\lambda)$ vs $q$: (a) for mobility $L_{ij}=\phi_i\delta_{ij}$, corresponding to Fig.~\ref{fig:fig2_firstorder_symmetriceps}(a); (b) for mobility $L_{ij}=\phi_i\delta_{ij}-\phi_i\phi_j$, corresponding to Fig.~\ref{fig:figA1}. Homogeneous state composition $(\phi_1^0,\phi_2^0)=(0.12,0.45)$ in both cases. The different mobility is seen to lead to a dominant static spinodal mode in (a) but an oscillatory one in (b).}
        \label{fig:figA2}
    \end{figure}

A comparison between Figs.~\ref{fig:fig2_firstorder_symmetriceps}(a) and~\ref{fig:figA1} shows that, as expected from the $q\to 0$ analysis in Fig.~\ref{fig:fig1_mobility_change}, the form of the mobility in non-reciprocal Model B can affect whether the spinodal dynamics is dominated by static or oscillatory modes. As an example we show in Fig.~\ref{fig:figA2}, for the location $(\phi_1^0,\phi_2^0)=(0.12,0.45)$ in the spinodal phase diagrams of Figs.~\ref{fig:fig2_firstorder_symmetriceps}(a) and~\ref{fig:figA1}, respectively, the corresponding rate matrix spectra. These plots demonstrate that the dominant spinodal mode is static in Fig.~\ref{fig:figA2}(a), but oscillatory in Fig.~\ref{fig:figA2}(b), with the only difference between the two situations being the mobility.

\subsection{Numerical simulations with the two mobilities}

\begin{figure}
        \centering
\includegraphics[width=\linewidth]{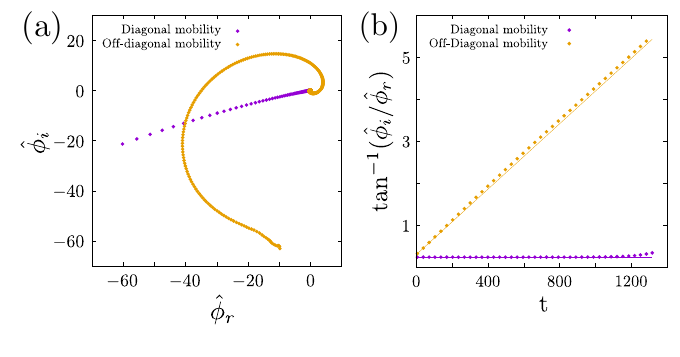}
\caption{(a) Real and imaginary part of (projected) density Fourier mode at wavevector $q_{\rm max}$ eigenvalue from simulations, for system parameters corresponding to Fig.~\ref{fig:figA2}(a) (diagonal mobility, $L_{ij}=\phi_i\delta_{ij}$) and Fig.~\ref{fig:figA2}(b) (off-diagonal mobility, 
$L_{ij}=\phi_i\delta_{ij}-\phi_i\phi_j$).}
        \label{fig:figA3}
    \end{figure}

We confirm the predictions from the linearized dynamics for the two systems in Fig.~\ref{fig:figA2} by direct numerical simulation of the full nonlinear non-reciprocal Model B equations for the parent composition $(\phi_1^0,\phi_2^0) = (0.12, 0.45)$. We project the Fourier transform of the density with wavevector $q_{\rm max}$ onto one of the two eigenvectors determined from the linearized dynamics as explained in App.~\ref{appendix: num_method_oscillatory_analysis}, and then obtain the real and imaginary parts $\hat{\phi}_r$ and $\hat{\phi}_i$. The results are shown in Fig~\ref{fig:figA3}.

One observes that for the diagonal mobility $L_{ij}=\phi_i\delta_{ij}$, where from Fig.~\ref{fig:figA2}(a) we expect a static dominant spinodal mode, the trajectory of $(\hat{\phi}_r,\hat{\phi}_i)$, corresponding to the complex plane of the (projected) density Fourier mode, follows a straight line in the spinodal regime. The corresponding phase angle (Fig.~\ref{fig:figA3}(b)) is therefore constant until nonlinear effects set in.

For the case of Fig.~\ref{fig:figA2}(b), on the other hand, the dominant spinodal mode is predicted to be oscillatory. Fig.~\ref{fig:figA3}(a) confirms this, with $(\hat{\phi}_r,\hat{\phi}_i)$ tracing out an exponential spiral. The corresponding phase angle (Fig.~\ref{fig:figA3}(b)) grows linearly in time, with a prefactor that agrees with the imaginary part of the rate matrix eigenvalue as it should (thin line).

\begin{figure}
    \centering
    \includegraphics[width=1\linewidth]{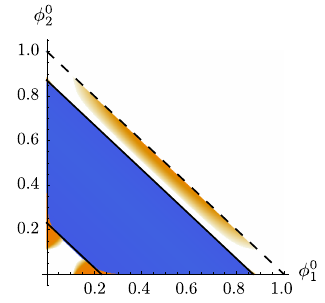}
    \caption{Phase diagram with highly non-reciprocal interface matrix: Blue region:- where we get complex eigenvalues with the above $\boldsymbol{K}$ matrix; black lines:- boundary of the region that we can expect to fill with high enough surface NR.}
    \label{fig:figA4}
\end{figure}

\section{Interfacial non-reciprocity}

\subsection{Spinodal phase diagram}\label{appendix:specific_interfacialNR_phasediagram}

In section \ref{results_surfNR}, we predict the maximal region in the spinodal phase diagram where we can get travelling spinodal waves in the presence of interfacial non-reciprocity. As a concrete example, we consider the set of highly nonreciprocal interfacial coefficients 
\begin{equation}
\boldsymbol{K} = \begin{pmatrix}
        0.34 & 0\\ -80 & 0.34
\end{pmatrix}
\label{nonreciproc_K}
\end{equation}
and otherwise choose the same system parameters as in Fig.~\ref{fig:fig1_mobility_change}(a). 
The resulting spinodal phase diagram is shown in Fig.~\ref{fig:figA4}. In accordance with our theoretical expectation, the region where spinodal travelling waves occur (blue) is significantly enlarged compared to 
Fig.~\ref{fig:fig1_mobility_change}(a). In fact it almost fills the maximal area (hatched in Fig.~\ref{fig:fig5_surfNR}(a), solid black lines in Fig.~\ref{fig:figA4}) where spinodal travelling waves can occur at all. 

   \begin{figure}[H]
    \centering
    \includegraphics[width=1\linewidth]{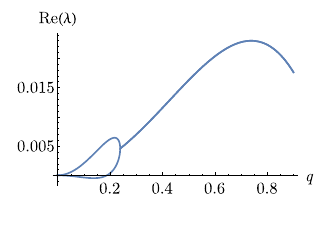}
\caption{Rate matrix spectrum $\text{Re}(\lambda)$ vs.\ $q$ for the system from Fig.~\ref{fig:figA4} at 
composition $(\phi_1^0,\phi_2^0)=(0.05,0.4)$.}
    \label{fig:figA6}
\end{figure}

\subsection{Numerical simulations}

\begin{figure*}
        \centering
        \includegraphics[width=1\linewidth]{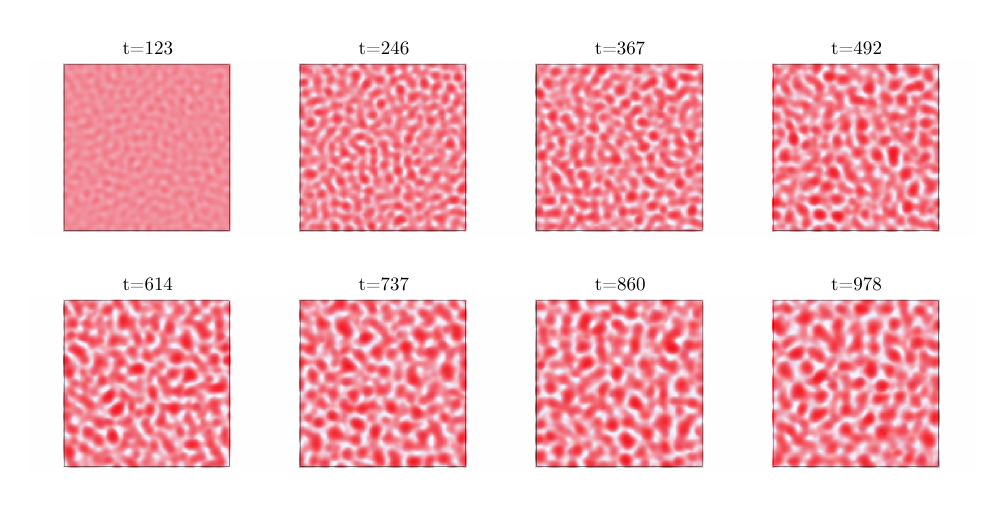}
        \caption{Snapshots of numerical simulations with interfacial non-reciprocity, for the system from Fig.~\ref{fig:figA4} with $(\phi_1^0,\phi_2^0)=(0.05,0.4)$. Travelling spinodal patterns are observed, as predicted.}
        \label{fig:figA5}
\end{figure*}

To illustrate the occurrence of travelling spinodal waves due to interfacial non-reciprocity, we perform numerical simulations at 
composition $(\phi_1^0,\phi_2^0)=(0.05,0.4)$. With interfacial coefficients $K_{ij}=\delta_{ij}$ as in Fig.~\ref{fig:fig2_firstorder_symmetriceps}, {\em i.e.}\ without interfacial non-reciprocity, the spinodal dynamics then produces static patterns ({\em cf.}\ Fig.~\ref{fig:fig2_firstorder_symmetriceps}(a)).

With interfacial non-reciprocity defined by~\eqref{nonreciproc_K}, however, we predict a travelling instability for the above composition, as can be seen from 
Fig.~\ref{fig:figA4} and the corresponding rate matrix spectrum 
in Fig.~\ref{fig:figA6}.
Careful inspection of the simulation snapshots in Fig.~\ref{fig:figA5} confirms that the spinodal patterns do indeed travel, as expected.
 
\section{Non-reciprocal Dean equation and coarse-graining} \label{appendix_deanNR}

An exact equation for dynamics of the density fields describing a mixture with non-reciprocal two-body interactions can be derived following Dean's original approach~\cite{David_S_Dean_1996}. We use in this appendix Greek letters $\alpha$, $\beta$ for the particle species, to have the Roman indices $i$, $j$ available for the particles.

Assuming overdamped dynamics, the Langevin equation for a particle $i$ of species $\alpha$ is
\begin{eqnarray}
    \gamma \dot{\boldsymbol{r}}_i = \sum_{\beta} \sum_{j\in \beta} \bm{F}_{\alpha\beta}(\boldsymbol{r}_{i}-\boldsymbol{r}_j)+\bm{\xi}_i
\end{eqnarray}
where $\bm{F}_{\alpha\beta}$ is the force exerted by a particle of species $\beta$ on a particle of species $\alpha$, $\gamma$ is the friction coefficient and $\bm{\xi}_i$ is thermal (Gaussian) white noise, with variance $2T\gamma$. With non-reciprocal interactions one has $\bm{F}_{\alpha\beta}\neq \bm{F}_{\beta\alpha}$. The density fields are defined as
\begin{eqnarray}
\phi_\alpha(\boldsymbol{r},t)=\sum_{i\in\alpha}w(\boldsymbol{r}-\boldsymbol{r}_i(t))
\end{eqnarray}
where $w(\boldsymbol{r})$ is a smearing function obeying $\int d\boldsymbol{r} \,w(\boldsymbol{r})=1$. Taking specifically the limit of no smearing, $w(\bm{r})=\delta(\bm{r})$, one then finds by following the steps in Ref.~\cite{David_S_Dean_1996} the equations of motion
\begin{eqnarray}   \frac{\partial\phi_\alpha(\boldsymbol{r},t)}{\partial t} = -\nabla\cdot\boldsymbol{j}_\alpha(\boldsymbol{r},t)
\label{Dean_exact}
\end{eqnarray}
with current
\begin{eqnarray}
\boldsymbol{j}_\alpha(\boldsymbol{r},t) &=& -D\nabla \phi_\alpha(\boldsymbol{r},t) \nonumber\\
&&{}+ \gamma^{-1}\phi_\alpha(\bm{r},t)\sum_\beta\int d\boldsymbol{r}' \bm{F}_{\alpha\beta}(\boldsymbol{r}-\boldsymbol{r}')\phi_\beta(\boldsymbol{r}',t)\nonumber\\
&&{}+ (2D\phi_\alpha(\boldsymbol{r},t))^{1/2}\bm{\eta}_\alpha(\boldsymbol{r},t)
\label{Dean_current}
\end{eqnarray}
where $D=T/\gamma$ and $\bm{\eta}_\alpha$ is unit variance Gaussian white noise. The second line contains the effect of the non-reciprocal interactions, via the density of $\phi_\alpha$ of species $\alpha$ times the velocity $\gamma^{-1}\int d\bm{r}'\,\bm{F}_{\alpha\beta}(\bm{r}-\bm{r}')\phi_\beta(\bm{r'},t)$ arising from the interactions.

While Eq.~\eqref{Dean_exact} is formally exact, for further analysis the $\phi_\alpha(\bm{r},t)$ are often treated as smooth density fields. Adopting this approximation and assuming that the interaction forces $\bm{F}_{\alpha\beta}(\bm{r})=-\nabla V_{\alpha\beta}(\bm{r})$ are derived from (non-reciprocal) potentials $V_{\alpha\beta}(\bm{r})$ that are central ($V_{\alpha\beta}(\bm{r})=V_{\alpha\beta}(|\bm{r}|)$), one finds within a gradient expansion to second order, {\em i.e.} 
\begin{equation}
\phi_\beta(\bm{r}')=
\phi_\beta(\bm{r}) + (\bm{r}'-\bm{r})\cdot\nabla\phi_\beta(\bm{r}) + 
\frac{1}{2}[(\bm{r}'-\bm{r})\cdot\nabla]^2\phi_\beta(\bm{r})
\end{equation}
for the second line of~\eqref{Dean_current}
\begin{eqnarray}
-\gamma^{-1}\phi_\alpha \sum_\beta
\nabla\left( \epsilon_{\alpha\beta}\phi_\beta - 
K_{\alpha\beta}\nabla^2 \phi_\beta
\right)
\end{eqnarray}
with
\begin{eqnarray}
\epsilon_{\alpha\beta} &=& \int d\bm{r} \,V_{\alpha\beta}(\bm{r})\\
K_{\alpha\beta}
&=& -\frac{1}{2d}\int d\bm{r}\,V_{\alpha\beta}(\bm{r})\,\bm{r}^2
\end{eqnarray}
where $d$ is the spatial dimension.
Inserting into~\eqref{Dean_exact} gives contributions to the equations of motion for the densities identical to those (up to the species index relabelling $\alpha,\beta \to i,j$) from the interaction chemical potentials~\eqref{mu_interface}, with mobility $L_{\alpha\beta}=\gamma^{-1}\phi_i\delta_{ij}$ in~\eqref{modB_current}. While this coarse-graining approach is clearly naive, given that {\em e.g.}\ it ignores the effects of short-range repulsion between particles and consequently becomes undefined for  interaction potentials $V_{\alpha\beta}(\bm{r})$ that diverge sufficiently strongly for $\bm{r}\to 0$, it does show that non-reciprocal interactions $V_{\alpha\beta}(\bm{r})$ will generically produce not only non-reciprocal bulk interaction coefficients $\epsilon_{\alpha\beta}$, but also non-reciprocal interfacial terms $K_{\alpha\beta}$. 

\bibliography{refs}

\end{document}